\documentclass{camera}
\usepackage{graphicx}
\usepackage{pstricks}
\usepackage{pst-coil}

\begin{document}
%
%---- SLASH
\def\slasha#1{#1\hskip-0.65em /}  %slasha per caratteri piccoli
\def\slashb#1{#1\hskip-1.3em /}   %slashb per quelli grandi
\def\slashc#1{#1\hskip-.4em /}
%
%---- UNITA` DI MISURA
\def \pb        {{\rm \, pb}}
\def \fb        {{\rm \, fb}}
\def \ipb       {{\rm \, pb^{-1}}}
\def \eV        {{\rm \,  eV}}
\def \keV       {{\rm \, keV}}
\def \MeV       {{\rm \, MeV}}
\def \GeV       {{\rm \, GeV}}
\def \TeV       {{\rm \, TeV}}
\def \TeVc      {\TeV/c}
\def \TeVcc     {\TeV/c^2}
\def \GeVc      {\GeV/c}
\def \GeVcc     {\GeV/c^2}
\def \MeVc      {\MeV/c}
\def \MeVcc     {\MeV/c^2}
%
%---- SIMBOLI
\def\ga{\mathrel{\raise.3ex\hbox{$>$\kern-.75em\lower1ex\hbox{$\sim$}}}}
\def\la{\mathrel{\raise.3ex\hbox{$<$\kern-.75em\lower1ex\hbox{$\sim$}}}}
\newcommand {\lesssim}
     {\,\raisebox{-0.6ex}{$\stackrel{\textstyle<}{\textstyle\sim}$}\,}
\newcommand {\gtrsim}
     {\,\raisebox{-0.6ex}{$\stackrel{\textstyle>}{\textstyle\sim}$}\,}
\newcommand{\ckm}{$\checkmark$}
%
%---- MISCELLANEA
\newcommand {\slashed}[1] { \mbox{\rlap{\hbox{/}} #1 }}
\newcommand {\onehalf}    {\raisebox{0.1ex}{${\frac{1}{2}}$}}
\newcommand {\fivethirds} {\raisebox{0.1ex}{${\frac{5}{3}}$}}
\newcommand {\OR}         {{\tt OR}\,}
\newcommand {\BR}         {{\rm BR}\,}
\newcommand {\rts}        {\sqrt{s}}
\newcommand {\lumi}       {\mathcal{L}}
\newcommand {\Lumi}       {\int\lumi\mathrm{d}t}
\newcommand {\gradi}    {^\circ}
\newcommand {\de}         {\partial}
\newcommand {\um}         {\; \mu \rm m}
\newcommand {\nm}         {\rm \; nm}
\newcommand {\us}         {\; \mu \rm s}
\newcommand {\cm}         {\rm \; cm}
\newcommand {\mm}         {\rm \; mm}
\newcommand {\m}          {\rm \; m}
\newcommand {\km}         {\rm \; km}
\newcommand {\V}          {\rm \; V}
\newcommand {\T}          {\rm \; T}
\newcommand {\kV}         {\rm \; kV}
\newcommand {\kVm}        {\rm \; kV\! / \! m} 
\newcommand {\MVm}        {\rm \; MV\! / \! m} 
\newcommand {\ns}         {\rm \; ns} 
%
%---- THEORY groups & AOB
\newcommand {\gws}        {\mathrm{SU(2)_L \otimes U(1)_Y}}
\newcommand {\sul}        {\mathrm{SU(2)_L}}
\newcommand {\suc}        {\mathrm{SU(3)_C}}
\newcommand {\ul}         {\mathrm{U(1)_Y}}
\newcommand {\uem}        {\mathrm{U(1)_{em}}}
\newcommand {\sigmabar}   {\overline{\sigma}}
\newcommand {\gmunu}      {g^{\mu \nu}}
\newcommand {\munu}       {{\mu \nu}}
\newcommand {\obra}       {\langle 0 |}
\newcommand {\oket}       {| 0 \rangle}
%
%---- THEORY lepton fields
\newcommand {\LL}         {L^{\alpha}_{\mathrm L}}
\newcommand {\LLd}        {L^{\dagger \alpha}_{\mathrm L}}
\newcommand {\lL}         {\ell^{\alpha}_{\mathrm L}}
\newcommand {\lLd}        {\ell^{\dagger \alpha}_{\mathrm L}}
\newcommand {\ld}         {\ell^{\dagger \alpha}}
\newcommand {\lb}         {\overline{\ell}^{\alpha}}
\newcommand {\lR}         {\ell^{\alpha}_{\mathrm R}}
\newcommand {\lRd}        {\ell^{\dagger \alpha}_{\mathrm R}}
\newcommand {\nuL}        {\nu^{\alpha}_{\mathrm L}}
\newcommand {\nuLb}       {\overline{\nu}^{\alpha}_{\mathrm L}}
\newcommand {\nub}        {\overline{\nu}^{\alpha}}
\newcommand {\lept}       {\ell^\alpha}
\newcommand {\neut}       {\nu^{\alpha}}
\newcommand {\nuLd}       {\nu^{\dagger \alpha}_{\mathrm L}}
\newcommand {\Phid}       {\Phi^\dagger}
%
%---- THEORY quark fields
\newcommand {\up}         {u^{\alpha}}
\newcommand {\ub}         {\overline{u}^{\alpha}}
\newcommand {\down}       {d^{\alpha}}
\newcommand {\db}         {\overline{d}^{\alpha}}
\newcommand {\QL}         {Q^{\alpha}_{\mathrm L}}
\newcommand {\QLd}        {Q^{\dagger \alpha}_{\mathrm L}}
\newcommand {\UL}         {U^{\alpha}_{\mathrm L}}
\newcommand {\ULd}        {U^{\dagger \alpha}_{\mathrm L}}
\newcommand {\UR}         {U^{\alpha}_{\mathrm R}}
\newcommand {\URd}        {U^{\dagger \alpha}_{\mathrm R}}
\newcommand {\DL}         {D^{\alpha}_{\mathrm L}}
\newcommand {\DLd}        {D^{\dagger \alpha}_{\mathrm L}}
\newcommand {\DR}         {D^{\alpha}_{\mathrm R}}
\newcommand {\DRd}        {D^{\dagger \alpha}_{\mathrm R}}
\newcommand {\bfell}      {\ell\kern-0.4em
                           \ell\kern-0.4em
                           \ell\kern-0.4em
                           \ell }
\newcommand {\obfell}     {\overline{\ell}\kern-0.4em
                           \overline{\ell}\kern-0.4em
                           \overline{\ell}\kern-0.4em
                           \overline{\ell}}
\newcommand {\bfH}      {\; {\cal H}\kern-0.5em \kern-0.4em
                           {\cal H}\kern-0.5em \kern-0.4em
                           {\cal H}\kern0.1em }
\newcommand {\obfH}     {\; \overline{\cal H}\kern-0.5em \kern-0.4em 
                           \overline{\cal H}\kern-0.5em \kern-0.4em 
                           \overline{\cal H}\kern0.1em }
%
%---- PARTICELLE
\def \b             {{\mathrm b}}
\def \t             {{\mathrm t}}
\def \charm         {{\mathrm c}}
\def \d             {{\mathrm d}}
\def \u             {{\mathrm u}}
\def \e             {{\mathrm e}}
\def \q             {{\mathrm q}}
\def \g             {{\mathrm g}}
\def \p             {{\mathrm p}}
\def \s             {{\mathrm s}}
\def \n             {{\mathrm n}}
\def \l             {\ell} 
\def \f             {{f}} 
\def \D             {{\mathrm D}}
\def \K             {{\mathrm K}}
\def \Z             {{\mathrm Z}}
\def \W             {{\mathrm W}}
\def \S             {{\mathrm S}}
\def \N             {{\mathrm N}}
\def \L             {{\mathrm L}}
\def \R             {{\mathrm R}}
\def \P             {{\mathrm P}}
\def \G             {{\mathrm G}}
%
%---- SUSY
\newcommand {\dm}         {\Delta m}
\newcommand {\dM}         {\Delta M}
\newcommand {\ldm}        {\mbox{``low $\dm$''}}
\newcommand {\hdm}        {\mbox{``high $\dm$''}}
\newcommand {\nnc}        {{\overline{\mathrm N}_{95}}}
\newcommand {\snc}        {{\overline{\sigma}_{95}}}
\newcommand {\susy}       {{supersymmetry}}
\newcommand {\susyc}      {{supersymmetric}}
\newcommand {\aj}         {\mbox{\sf AJ}}
\newcommand {\ajl}        {\mbox{\sf AJL}}
\newcommand {\llh}        {\mbox{\sf LLH}}
%
%---- SPARTICELLE
\newcommand {\rpc}     {{\rm RPC}}
\newcommand {\rpv}     {{\rm RPV}}
\newcommand {\sfe}     {{\tilde{f}}}
\newcommand {\sfL}     {{\tilde{f}_{\mathrm L}}}
\newcommand {\sfR}     {{\tilde{f}_{\mathrm R}}}
\newcommand {\sfone}   {{\tilde{f}_{1}}}
\newcommand {\sftwo}   {{\tilde{f}_{2}}}
\newcommand {\sneu}    {{\tilde{\nu}}}
\newcommand {\wino}    {{\mathrm{\widetilde{W}}}}
\newcommand {\bino}    {{\mathrm{\widetilde{B}}}}
\newcommand {\se}      {{\mathrm{\tilde{e}}}}
\newcommand {\seR}     {{\mathrm{\tilde{e}_{R}}}}
\newcommand {\seL}     {{\mathrm{\tilde{e}_{L}}}}
\newcommand {\st}      {{\mathrm{\tilde{\tau}}}}
\newcommand {\stR}     {{\mathrm{\tilde{\tau}_{R}}}}
\newcommand {\stL}     {{\mathrm{\tilde{\tau}_{L}}}}
\newcommand {\stone}   {{\mathrm{\tilde{\tau}_{1}}}}
\newcommand {\sttwo}   {{\mathrm{\tilde{\tau}_{2}}}}
\newcommand {\sm}      {{\mathrm{\tilde{\mu}}}}
\newcommand {\smR}     {{\mathrm{\tilde{\mu}_{R}}}}
\newcommand {\smL}     {{\mathrm{\tilde{\mu}_{L}}}}
\newcommand {\Sup}     {{\mathrm{\tilde{u}}}}
\newcommand {\suR}     {{\mathrm{\tilde{u}_{R}}}}
\newcommand {\suL}     {{\mathrm{\tilde{u}_{L}}}}
\newcommand {\sdo}     {{\mathrm{\tilde{d}}}}
\newcommand {\sdR}     {{\mathrm{\tilde{d}_{R}}}}
\newcommand {\sdL}     {{\mathrm{\tilde{d}_{L}}}}
\newcommand {\sch}     {{\mathrm{\tilde{c}}}}
\newcommand {\scR}     {{\mathrm{\tilde{c}_{R}}}}
\newcommand {\scL}     {{\mathrm{\tilde{c}_{L}}}}
\newcommand {\sst}     {{\mathrm{\tilde{s}}}}
\newcommand {\ssR}     {{\mathrm{\tilde{s}_{R}}}}
\newcommand {\ssL}     {{\mathrm{\tilde{s}_{L}}}}
\newcommand {\stopR}   {{\tilde{\mathrm{t}}_{R}}}
\newcommand {\stopL}   {{\tilde{\mathrm{t}}_{L}}}
\newcommand {\stopone} {{\tilde{\mathrm{t}}_{1}}}
\newcommand {\stoptwo} {{\mathrm{\tilde{t}_{2}}}}
\newcommand {\sto}     {{\tilde{\mathrm{t}}}}
\newcommand {\SQ}      {{\mathrm{\widetilde{Q}}}}
\newcommand {\STO}     {{\mathrm{\widetilde{T}}}}
\newcommand {\glu}     {{\mathrm{\tilde{g}}}}
\newcommand {\sbotR}   {{\mathrm{\tilde{b}_{R}}}}
\newcommand {\sbotL}   {{\mathrm{\tilde{b}_{L}}}}
\newcommand {\sbotone} {{\mathrm{\tilde{b}_{1}}}}
\newcommand {\sbottwo} {{\mathrm{\tilde{b}_{2}}}}
\newcommand {\sbot}    {{\tilde{\mathrm{b}}}}
\newcommand {\squa}    {{\tilde{\mathrm{q}}}}
\newcommand {\squal}   {{\tilde{\mathrm{q}}_{\rm L}}}
\newcommand {\squar}   {{\tilde{\mathrm{q}}_{\rm R}}}
\newcommand {\sqL}     {{\tilde{\mathrm{q}}_{\rm L}}}
\newcommand {\sqR}     {{\tilde{\mathrm{q}}_{\rm R}}}
\newcommand {\snu}     {{\tilde{\nu}}}
\newcommand {\snue}    {{\tilde{\nu}_{\mathrm e}}}
\newcommand {\snum}    {{\tilde{\nu}_{\mu}}}
\newcommand {\snut}    {{\tilde{\nu}_{\tau}}}
\newcommand {\neu}     {{\chi}}
\newcommand {\chap}    {{\chi^+}}
\newcommand {\cham}    {{\chi^-}}
\newcommand {\chapm}   {{\chi^\pm}}

%
%---- SUSY PARAMETRI
\newcommand {\thstop} {\mathrm{\theta_{\tilde{t}}}}
\newcommand {\thsbot} {\mathrm{\theta_{\tilde{b}}}}
\newcommand {\thsqua} {\mathrm{\theta_{\tilde{q}}}}
\newcommand {\Mcha}{M_{\chi^\pm}}
\newcommand {\Mchi}{M_\chi}
\newcommand {\Msnu}{M_{\tilde{\nu}}}
\newcommand {\tanb}{\tan\beta}
%
%---- ABBREVIAZIONI

%
%---- PROCESSI FISICI
\newcommand {\rb}    {{\rm R_{\b}}}
\newcommand {\qq}    {{\q \overline{\q}}}
\newcommand {\bb}    {{\b \overline{\b}}}
\newcommand {\ff}    {{\f \bar{\f}}}
\newcommand {\el}    {{\e ^+}}
\newcommand {\po}    {{\e ^-}}
\newcommand {\ee}    {{\e ^+ \e ^-}}
\newcommand {\fbody} {{\sto \to \b \chi {\rm f \bar{f}'}}}
\newcommand {\gaga}  {\gamma\gamma}
\newcommand {\ggqq}  {\gamma\gamma \rightarrow \q\overline{\q}}
\newcommand {\ggtt}  {\gamma\gamma \rightarrow \tau^{+}\tau^{-}}
\newcommand {\qqg}   {\q\overline{\q}\gamma}
\newcommand {\ttg}   {\tau^{+}\tau^{-}\gamma}
\newcommand {\wenu}  {{\rm We\nu_\e}}
\newcommand {\gsZ}   {\gamma^\star\mathrm{Z}}
\newcommand {\ggh}   {\gamma\gamma\rightarrow{\mathrm{hadrons}}}
\newcommand {\ZZg}   {\mathrm ZZ^{*}/\gamma^{*}}
%
%---- VARIABILI
\newcommand {\zo}      {{z_0}}
\newcommand {\ip}      {{d_0}}
\newcommand {\thr}     {{{\rm thrust}}}
\newcommand {\athr}    {{\hat{\rm a}_{\rm thrust}}}
\newcommand {\ththr}   {{\theta_{\rm thrust}}}
\newcommand {\acol}    {{\Phi_{\rm acol}}}
\newcommand {\acop}    {{\Phi_{\rm acop}}}
\newcommand {\acopt}   {{\Phi_{\rm acop_T}}}
\newcommand {\thpoint} {\theta_{\rm point}}
\newcommand {\thscat}  {\theta_{\rm scat}}
\newcommand {\etwelve} {E_{12\gradi}}
\newcommand {\ethirty} {E_{30\gradi}}
\newcommand {\eiso}[1] {E^{\, \triangleleft 30\gradi}_{#1}}
\newcommand {\phimiss} {{\phi_{\vec{p}_{\rm miss}}}}
\newcommand {\ewedge}  {E(\phi_{\vec{p}_{\rm miss}}\pm 15\gradi)}
\newcommand {\evis}    {E_{\rm vis}}
\newcommand {\etot}    {E_{\rm vis}}
\newcommand {\emis}    {E_{\rm miss}}
\newcommand {\mvis}    {M_{\rm vis}}
\newcommand {\mtot}    {M_{\rm vis}}
\newcommand {\mmis}    {M_{\rm miss}}
\newcommand {\mhad}    {M^{\rm ex \, \ell_1}_{\rm vis}}
\newcommand {\mhadtwo} {M^{\rm ex \, \ell_1\ell_2}_{\rm vis}}
\newcommand {\ehad}    {E^{\rm NH}_{\rm vis}}
\newcommand {\epho}    {E^{\gamma}_{\rm vis}}
\newcommand {\echa}    {E^{\rm ch}_{\rm vis}}
\newcommand {\nch}     {{N_{\rm ch}}}
\newcommand {\elept}   {E_{\rm lept}}
\newcommand {\elepone} {E_{\ell _1}}
\newcommand {\eleptwo} {E_{\ell _2}}
\newcommand {\pvis}    {{\vec{p}_{\rm vis}}}
\newcommand {\pmis}    {{\vec{p}_{\rm miss}}}
\newcommand {\thmiss}  {{\theta_{\pmis}}}
\newcommand {\pt}      {{p_{\rm t}}}
\newcommand {\ptch}    {{p_{\rm t}^{\rm ch}}}
\newcommand {\pch}    {{p^{\rm ch}}}
\newcommand {\pz}      {{p_z}}
\newcommand {\ptnoNH}  {{p_{\rm t}^{\rm ex \, NH}}}
\newcommand {\puds}    {{P_{\rm uds}}}
%
%
% no more of Christian's random capitalization!
% more of mine
\newcommand{\brchal}{\cal{B}($\PCha \rightarrow \ell\nu\PChi\ $)}
\newcommand{\M}{M_{2}}
\newcommand{\Mp}{M_{2}}
\newcommand{\sigbg}{\sigma_{\mathrm{bg}}}
\newcommand{\ww}   {\mathrm {WW}}
\newcommand{\zz}   {\mathrm Z\gamma^{*}}
\newcommand{\ewnu} {\mathrm{eW}\nu}
\newcommand{\eez}  {\mathrm {eeZ}}
\newcommand{\gagall}{{\gamma\gamma\rightarrow \ell\ell }}
\newcommand{\Pstaup}{{\widetilde{\tau}_{1}}}
\newcommand{\Pstaul}{{\widetilde{\tau}_{L}}}
\newcommand{\Pstaur}{{\widetilde{\tau}_{R}}}
\newcommand{\mzero}{m_{0}}
\newcommand{\msnu}{M_{\tilde{\nu}}}
\newcommand{\mcha}{M_{\chi^{\pm}}}
\newcommand{\mchi}{M_{\chi}}
\newcommand{\mstau}{M_{{\widetilde{\tau}_{1}}}}
\newcommand{\atau}{A_{\tau}}
\newcommand{\chsnu}{\PCha \rightarrow \ell \tilde{\nu}}
\newcommand{\chstau}{\PCha \rightarrow \tilde{\tau}_{1}\nu}
\newcommand{\chlep}{\PCha \rightarrow \ell\nu\chi}
\newcommand{\Tcsq}{\mathrm{TeV}/c^2}
% new for thesis
\newcommand{\nobs}{N_{\mathrm{obs}}}
\newcommand{\nlim}{N_{\mathrm{lim}}}
\newcommand{\Brl}{\cal{B}_{\ell}}
\newcommand{\leff} {\mathcal{L}_{\mathrm{eff}}}
\newcommand{\dedx}{{\mathrm{d}}E/{\mathrm{d}}x}
\newcommand{\chtau}{\PCha \rightarrow \tau\nu\chi}
\newcommand{\ssqtw}{\sin^{2}\theta_{\mathrm W}}
\newcommand{\nnz}{{\mathrm \nu\bar{\nu}Z}}
% added by bill
\def \ggll    {\gamma\gamma \rightarrow \ell^{+}{\ell}^{-}}
\def \tautau  {\mathrm \tau^{+}\tau^{-}}
\def \ffg  {f\bar{f}(\gamma)}
\def \lll   {\ell^{+}{\ell}^{-}}
\def \ww   {\mathrm WW}
\def \zz   {\mathrm Z\gamma^{*}}
\def \znn  {\mathrm Z\nu\nu}
\def \zee  {\mathrm Zee}
\def \rts  {\sqrt{s}}
\def \mstop {m_{\tilde{\mathrm{t}}}}
\def \msnu  {m_{\tilde{\nu}}}
\def \elow   {E_{12^{\circ}}}
\def \gev    { \, \mathrm{GeV}/\it{c}^{\mathrm{2}}}
\def \gvm    { \, \mathrm{GeV}/\it{c}}
\def \mx     {M_{\mathrm{eff}}} 
\newcommand{\neutr}{\chi}
%end fabio

%dalla mia pretesi

%\def \X             {\mathrm X} 
%\def \V             {\mathrm V} 
\def \Zcc           {\Z \to \charm \bar{\charm} }
\def \Zbb           {\Z \to \b \bar{\b} }
\def \decDS         {\D^{*+} \to \D^0 \pi^+}
\def \decsDS        {\D^{*+} \to \D^0 \pi^+_s}
\def \deckp         {\D^{0} \to \K^- \pi^+}
\def \deckppp       {\D^{0} \to \K^- \pi^+ \pi^+ \pi^-}
\def \deckpp        {\D^{0} \to \K^- \pi^+ \pi^0}
\def \deckpS        {\D^{0} \to \K^- \pi^+ (\pi^0)}
\def \decskp        {\D^{*+} \to \pi^{+}_{s} \K^- \pi^+}
\def \decskppp      {\D^{*+} \to \pi^{+}_{s} \K^- \pi^+ \pi^+ \pi^-}
\def \decskpp       {\D^{*+} \to \pi^{+}_{s} \K^- \pi^+ \pi^0}
\def \decskpS       {\D^{*+} \to \pi^{+}_{s} \K^- \pi^+ (\pi^0)}
\def \epsc          {\varepsilon_{\charm}}
\def \epsb          {\varepsilon_{\b}}
\def \pctod         {P_{\charm \to \D^*}}
\def \pbtod         {P_{\b \to \D^*}}
\def \Gbb           {\Gamma_{\b\bar{\b}}}
\def \Gcc           {\Gamma_{\charm\bar{\charm}}}
\def \Gh            {\Gamma_{\mathrm h}}

\title{THE TOPOLOGIES OF\\ SUPERSYMMETRY SIGNALS AT LEP}

%%%%%%%%%%%%%%%%%%%%%%%%%%%%%%%%%%%%%%%%%%%%%%%%%%%%%%%%
%
\author{Giacomo Sguazzoni}

%%%%%%%%%%%%%%%%%%%%%%%%%%%%%%%%%%%%%%%%%%%%%%%%%%%%%%%%
%
\organization{European Laboratory for Particle Physics (CERN)\\
CH-1211 Geneva 23, Switzerland} 

\maketitle

{\small 
\begin{center}
\parbox[h][16ex][t]{.8\textwidth}
{\begin{center}
{\bf Abstract}
\end{center}
\small The topologies arising from the production of supersymmetric
particles at the LEP collider are briefly reviewed recalling detector
requirements, simulation and other experimental issues.} 
\end{center}
}

%%%%%%%%%%%%%%%%%%%%%%%%%%%%%%%%%%%%%%%%%%%%%%%%%%%%%%%%
% Write the text starting from here and using the usual
% LaTeX commands.
%
\section{Introduction}

Two years after the dismantling of LEP, most of the results of the
dedicated searches for production of supersymmetric particles
have been published or are close to publication~\cite{mazzuca}. The
analysed topologies are here briefly reviewed focusing on the
experimental issues faced by SUSY hunters. The huge experience gained
on this field is an important part of the LEP legacy. Hopefully, it
will turn out to be very useful for SUSY searches at the upcoming experiments.   

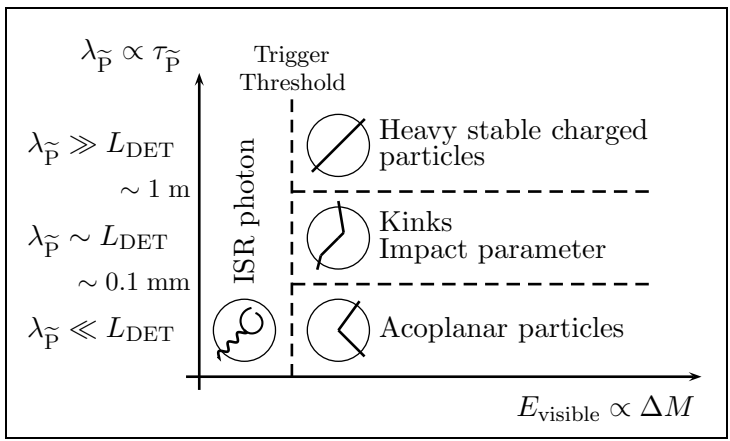
\begin{figure}[t]
\begin{center}
  \fbox{\parbox[t][5.5cm][b]{.74\textwidth}{
    \begin{picture}(0,0)
        \put(190,5){\mbox{$E_\mathrm{visible}\propto\Delta M$}}
        \put(25,140){\mbox{$\lambda_{\widetilde{\P}}\propto\tau_{\widetilde{\P}}$}}
        \put(138,110){\mbox{Heavy stable charged}}
        \put(138,100){\mbox{particles}}
        \put(138,75){\mbox{Kinks}}
        \put(138,65){\mbox{Impact parameter}}
        \put(138,35){\mbox{Acoplanar particles}}
        \put(5,105){\mbox{$\lambda_{\widetilde{\P}}\gg L_\mathrm{DET}$}}
        \put(40,88){\mbox{\small $\sim 1\m$}}
        \put(5,70){\mbox{$\lambda_{\widetilde{\P}}\sim L_\mathrm{DET}$}}
        \put(24,53){\mbox{\small $\sim 0.1\mm$}}
        \put(5,35){\mbox{$\lambda_{\widetilde{\P}}\ll L_\mathrm{DET}$}}
        \put(83,55){\rotatebox{90}{\mbox{ISR photon}}}
        \put(20,0){\psline[linewidth=1pt]{->}(45pt,20pt)(240pt,20pt)}
        \put(20,0){\psline[linewidth=1pt]{->}(50pt,15pt)(50pt,135pt)}
        \put(105,90){\psline[linewidth=1pt, linestyle=dashed]{-}(0pt,0pt)(135pt,0pt)}
        \put(105,55){\psline[linewidth=1pt, linestyle=dashed]{-}(0pt,0pt)(135pt,0pt)}
        \put(105,20){\psline[linewidth=1pt, linestyle=dashed]{-}(0pt,0pt)(0pt,105pt)}
%
%Acop
	\put(119.,37.5){
	\pscircle[linewidth=.5pt](0,0){12pt}
	\psline[linewidth=1pt](0,0)(8.pt,11.pt)
	\psline[linewidth=1pt](0,0)(9.9pt,-9.9pt)
	}
%
%IP/Kinks
	\put(119.,72.5){
	\pscircle[linewidth=.5pt](0,0){12pt}
	\psline[linewidth=1pt](0,0)(2pt,2pt)
	\psline[linewidth=1pt](2pt,2pt)(0pt,14pt)
	\psline[linewidth=1pt](0,0)(-6.6pt,-6.6pt)
	\psline[linewidth=1pt](-6.6pt,-6.6pt)(-8pt,-12pt)
	}
%
%HSCP
	\put(119.,107.5){
	\pscircle[linewidth=.5pt](0,0){12pt}
	\psline[linewidth=1pt](-10pt,-10pt)(10pt,10pt)
	}
%
%ISR
	\put(83.5,37.5){
	\pscircle[linewidth=.5pt](0,0){12pt}
	\pscoil[coilarm=0,coilaspect=0,coilheight=2,coilwidth=2.5pt,linewidth=1pt](0,0)(-10pt,-10pt)
	\psarc(3.53pt,3.53pt){5pt}{100}{230}
	\psarc(2.8pt,2.8pt){4pt}{230}{70}
	}
        \put(85,145){{\parbox[t][1cm][b]{40pt}{
	\begin{center}
	{\footnotesize
	Trigger\\ 
	\vskip -3pt
	Threshold
	}
	\end{center}
	}}}
    \end{picture}
    }}
    \caption{A schematic diagram of the $\lambda_{\widetilde{\P}}$
	vs. $E_\mathrm{visible}$ plane reporting the relevant topologies
	within the R-parity conserving scenario.}
	\label{fig:topo}
\end{center}
\end{figure}    

\section{Overview of topologies}

R-parity conservation is assumed: the LSP (Lightest Supersymmetric 
Particle) is stable, and, most probably, also neutral and weakly
interacting; at LEP the sparticles are pair produced; the decay
processes bring to final states containing at least one LSP.

The variety of supersymmetric models leads to a very wide
phe\-no\-me\-no\-lo\-gy. For each type of final state (hadronic
jets, leptons, $\gamma$'s) many topologies exist, depending on
$E_\mathrm{visible}$, the energy of the visible system, and on
$\lambda_{\widetilde{\P}}$, the sparticle decay length. At a given
collision energy, $E_\mathrm{visible}$ is proportional to the mass
difference between the sparticle and the escaping LSP ($\Delta M =
M_{\widetilde{\P}} - M_\mathrm{LSP}$) and $\lambda_{\widetilde{\P}}$ 
is related to the couplings and to $\Delta M$ by means of the
sparticle lifetime $\tau_{\widetilde{\P}}$.

An exhaustive search for SUSY at LEP has been pursued by using several
different analyses, which, however, can be roughly grouped by experimental
topologies, as illustrated in Figure~\ref{fig:topo}:\\
$\bullet$ energy of the visible system above the detector sensitivity
($\sim \GeV$ trigger threshold): {\em acoplanar particles}, {\em
impact parameter or kinks} and {\em heavy stable charged particles},
respectively for $\lambda_{\widetilde{\P}}$ smaller, comparable or
greater than the typical detector dimension;\\  
$\bullet$ energy of the visible system below the detector sensitivity:
{\em ISR photon}, the hard initial state radiation photon being used for trigger.

The analyses for the R-parity violating scenario, also widely faced
at LEP, search for a different set of topologies, not discussed here.

{\bf Acoplanar particles}. The main signature of this topology,
arising in case of pair production of sparticles decaying promptly, is
the missing energy and momentum from escaping LSP's, hence the
hermeticity is the main detector requirement. The selections often
implement fiducial cuts to avoid missing energy and momentum to be
faked by cracks or dead zones in calorimeters and luminosity monitors.

Signal event shape and backgrounds, and thus selections, heavily
depend on $\Delta M$.  The small $\Delta M$ region is the most
problematic since the visible system is soft and the dominant background,
$\ee\to\gaga$, has a huge cross section ($\gtrsim 10$nb) to be compared with
the typical SUSY cross sections ($\sim 0.1$pb). The strong $\gaga$
rejection factor required ($\sim 10^5$) results in degraded
signal efficiencies. Large efforts have been spent to get the background 
estimation reliable despite the large samples needed and the complex
physics underlying the elementary $\gaga$ processes.

Powerful anti-$\gaga$ criteria exist; requiring small energy
deposition in the forward directions ($|\theta|\lesssim10^\circ$) is
one example. Unfortunately, the beam pipe and other
passive materials (as radiation shields) make this region 
intrinsically not hermetic, complicated in geometry and affected by
beam-related noise and backgrounds so that the evaluation of the detector
response is often difficult. These issues are taken into account within
the systematic uncertainties or overcome by applying cuts to safely
over-reject $\gaga$'s.

\begin{figure}[t]
  \begin{center}
    \includegraphics[width=0.47\textwidth, clip]{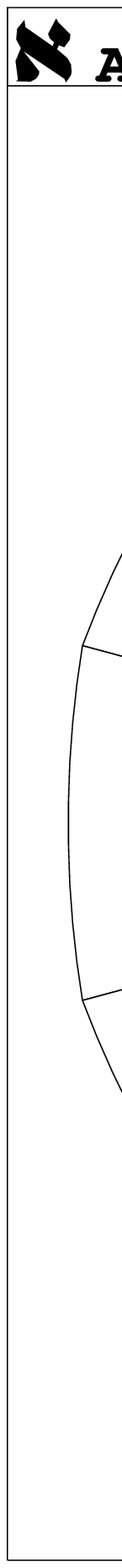}
    \hfill
    \includegraphics[width=0.47\textwidth, clip]{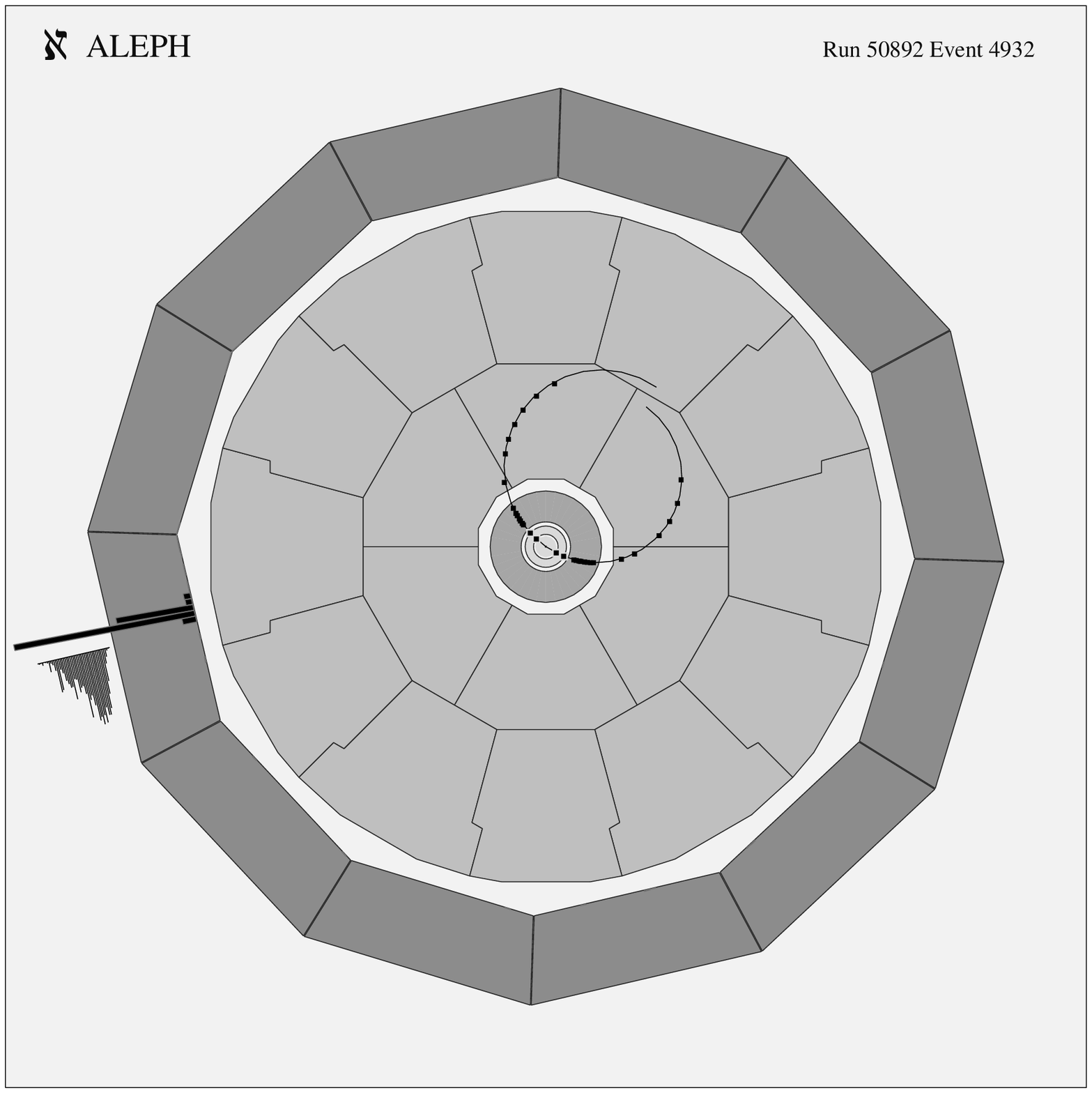}
    \begin{picture}(0,0)
      \put(-359,7){\mbox{(a)}}
      \put(-168,7){\mbox{(b)}}
    \end{picture}
    \caption{(a) Simulated event with long-lived sparticle
    pair leading to a kink and a track with large impact 
    parameter. (b) Very small $\Delta M$ sparticle signal 
    candidate with the triggering hard ISR photon.}
    \label{fig:IPkinks}
    \label{fig:isr}
  \end{center}
\end{figure}
{\bf Impact parameter or kinks}. This peculiar signature, illustrated
by the simulated event in Figure~\ref{fig:IPkinks}(a), can be
identified at LEP by means of the powerful tracking, even if the
reconstruction algorithms had in some cases to be suitably modified.
The non-negligible source of background due to material effects
(splash-backs and nuclear interactions), for which simulations are
normally not accurate enough, is kept under control by using topological
cuts or by requiring low vertex multiplicity.  
Cosmic rays recorded as $\ee$ interaction events can also
fake tracks not coming from the primary vertex. Good timing capability
are usually useful to reject these accidentals.

{\bf Heavy stable charged particles}. 
Heavy stable charged particles, produced with relatively low momentum,
can be easily tagged by means of particle identification capabilities. Most commonly 
dE/dx measurements are used, profiting of the large specific ionization
loss.

In case of heavy stable hadrons, a big issue is represented 
by the si\-mu\-la\-tion of the aspects related to the hadronization and to the
strong interaction into the calorimeters. The simulations are based
on some reasonable extrapolation of the behaviour of known particles;
however, as an extra safety margin, the selections have been designed
to make minimal use of the calorimetric informations. 

{\bf ISR photon}. If $\Delta M$ falls below few hundreds of MeV's,
the triggering on the sparticle signal relies only on hard photons 
from initial state radiation. As an example, Figure~\ref{fig:isr}(b)
reports a candidate event selected in such type of analyses. The
method allows to extend the sensitivity down to $\sim 150\MeV$;
reliable ISR simulations are crucial for a correct evaluation of the
resulting tiny efficiencies ($\lesssim 1\%$).

\section{Conclusion}
Despite the frustrating negative outcome, hunting for Supersymmetry
at LEP turned out to be a challenging search for topologies not
present or rare in the Standard Model. The experimental issues
required a stimulating and deep comprehension of the detectors and the
development of new techniques and dedicated algorithms.

\section{Acknowledgments}
I would like to thank P.~Azzurri and G.~Ganis for their valuable help and
many useful discussions.

\end{document}